\documentclass[iop]{emulateapj}
\usepackage{natbib}

\newcommand{\bmath}[1]{\mbox{\protect\boldmath$#1$}}

\shorttitle{Extremal energy shifts of radiation from a ring...}
\shortauthors{V.~Karas and V.~Sochora}

\begin{document}

\title{Extremal energy shifts of radiation from a ring near a rotating black hole}


\author{Vladim\'{\i}r Karas, and Vja\v{c}eslav Sochora}
\affil{Astronomical Institute, Academy of Sciences, Bo\v{c}n\'{\i}~II~1401, CZ-14131~Prague, Czech Republic}






\begin{abstract}
Radiation from a narrow circular ring shows a characteristic double-horn 
profile dominated by photons having energy around the maximum or minimum 
of the allowed range, i.e.\ near the extremal values of the energy
shift.  The energy span of a spectral line is a function of the ring
radius, black  hole spin, and observer's view angle. We describe a 
useful approach to calculate the extremal energy shifts in the regime
of strong gravity. Then we consider an accretion disk consisting of a
number of separate nested  annuli in the equatorial plane of Kerr black
hole, above the innermost stable circular orbit (ISCO). We suggest that
the radial structure of the disk emission could be reconstructed using
the extremal energy shifts of the individual rings deduced from the
broad wings of a relativistic spectral line.
\end{abstract}

\section{Introduction}
Emission from inner regions of accretion disks around black holes
provides wealth of information about matter in extreme conditions.
Relativistic spectral line of iron, broadened and skewed by fast orbital
motion and redshifted by strong gravitational field, has been used to
constrain the parameters of the black hole, both in active galactic
nuclei \citep[AGN;][]{fabian00,reynolds03,miller07} and Galactic black
holes \citep{miller02,mcclintock06}. During recent years, much
discussion has revolved round the question of how close to the innermost
stable circular orbit \citep[ISCO, also called the marginally stable
orbit, $r=r_{\rm ms}$;][]{mtw73} the accretion disk extends, whether the
line is produced all the way down to the inner edge, and if the emission
from the accretion disk can be approximated by a smooth radial profile. 

\citet{reynolds97} point out that there could be some non-negligible
contribution to the reflection line originating even below ISCO. This
idea has been put in a more specific context of magnetized accretion
flows, as discussed e.g.\ by \citet{beckwith08}. On the other hand,
\citet{reynolds08} explored the flow properties close to the black hole
and they demostrated that the presence of ISCO leaves a strong imprint
on the X-ray reflection spectrum of the accretion disk due to the rapid
increase in ionization parameter. Furthermore, \citet{martocchia02a}
conclude, on the basis of the X-ray iron line modeling in GRS 1915+105
microquasar, that the line production is limited to the region above the
ISCO. Also \citet{svoboda10} found a convincing case for a disk being
truncated rather far above the ISCO (this time in a Seyfert 1.5 galaxy
IRAS 05078+1626), whereas \citet{turner10} suggest that a persistent
$5.44$~keV feature exhibited by another Seyfert~1 AGN, NGC~4051, could
originate from a preferred radius of the order of a few ISCO.

Despite a simple prediction for the radial dependence of the disk 
emission provided by the standard accretion disk scenario
\citep{novikov73,page74}, a realistic emissivity of a spectral line 
is not well constrained. In this paper we suggest that the function of
radial emissivity could be deduced  if the line is produced in discrete
rings rather than a whole continuous range of radii. Such an  assumption
is in fact a very realistic one; in the end the smooth radial profile
will have to be replaced by a more complicated emissivity law, which
will reflect the mechanism generating the line in a patchy disk
structure, perhaps originating from episodic accretion events. The
formation of detached annuli has been seen also in some models of
strongly magnetized plasma disks \citep{coppi06}, where they can develop
a periodic structure in radius. Although there is still a long way to
prove that such structures could emerge in radiation spectra, it is a
real possibility that should be tested observationally.

The existence of ring structure could be revealed by future detailed
spectroscopy of the spectral line wings. To this end we develop a
rigorous method of calculating the expected energy range of a spectral
line, taking into account the effects of strong gravity on photons
proceeding from the disk to the observer. These photons may follow
complicated routes, but we assume that they do not cross the equatorial
plane of the black hole and are neither absorbed nor scattered by
environment outside the accretion disk. We give accurate extremal shifts
over a wide range of parameters.

The adopted setup is relevant for geometrically thin, planar accretion
disks. Needless to say, the method will require high energy resolution
together with a sufficient number of counts in the observed spectrum.
The former condition is achievable with X-ray calorimeters. The latter
one imposes a more serious limitation, however, bright Galactic black
holes seem to be appropriate sources. Even if an immediate application
of the idea is not possible at present, the calculation of the extremal
energy shifts is by itself an interesting addition and of practical use
in future.

The paper is organized as follows. In Section~\ref{sec:broadline} we
introduce the model of nested rings as a representation of the iron line
emissivity from such a radially structured disk. In
Section~\ref{sec:extremalshifts} we describe the method of calculating
the extremal energy shifts. We provide an iterative semi-analytical
solution for the extremal shifts, $g_{\rm{}max}$ and $g_{\rm{}min}$, as
functions of three parameters: the emission radius $r_{\rm{}em}$, spin
parameter $a$, and the inclination angle $i$. In order to demonstrate
the dependencies, we show a graphical representation of $g_{\rm{}max}$
and $g_{\rm{}min}$ in terms of parametric plots over the three-dimensional
parameter space. Finally, in sections~\ref{sec:discussion} and
\ref{sec:conclusions} we discuss and summarize our results.

\begin{figure*}[tbh!]
\includegraphics[width=.49\textwidth]{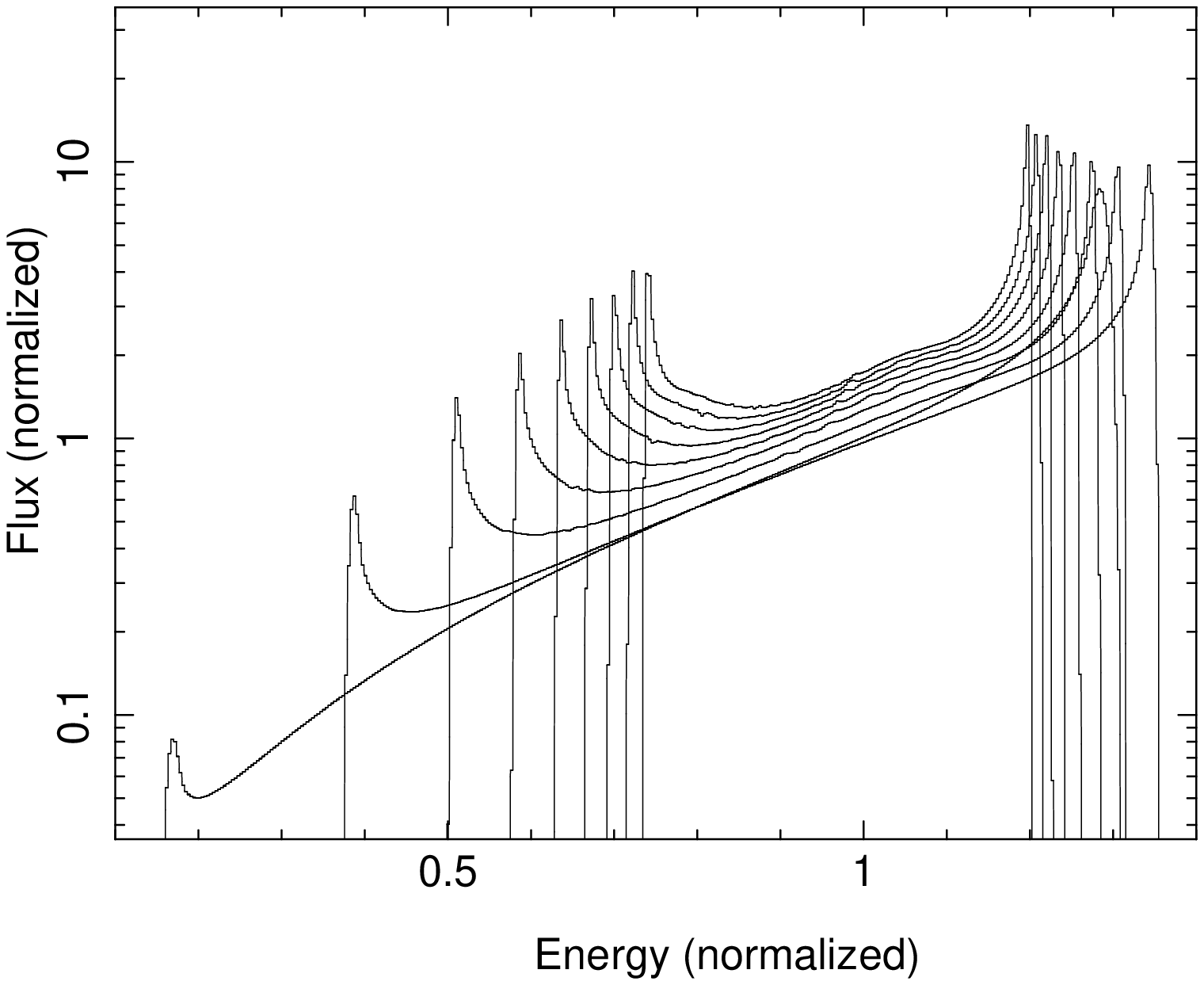}
\hfill
\includegraphics[width=.49\textwidth]{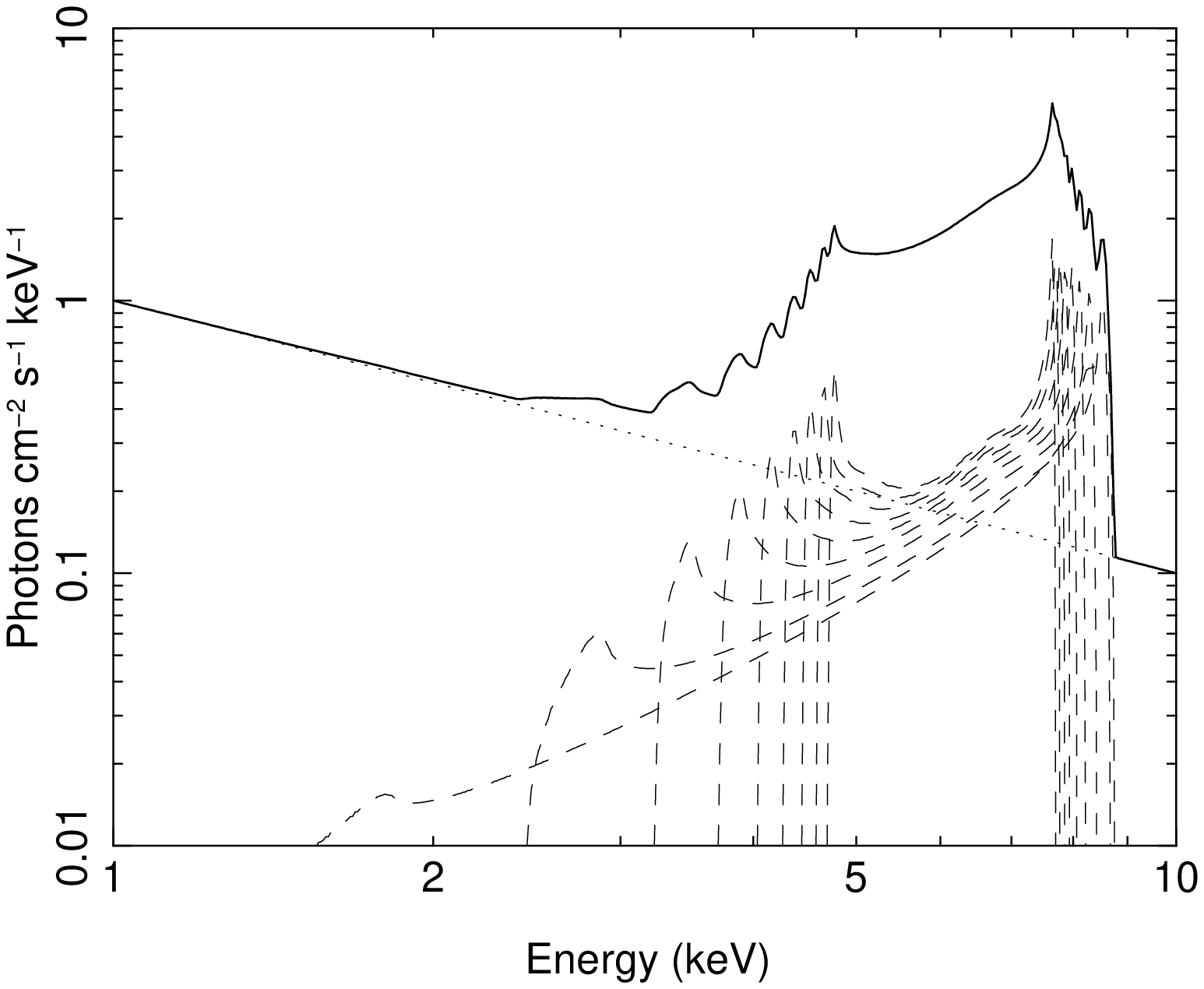}
\caption{Forming a double-horn spectral line by
superposing profiles of several narrow-rings. Left: theoretical 
profiles from a set of nine infinitesimally narrow rings orbiting in the
equatorial plane of a Kerr black hole. Radii of the rings increase
equidistantly from $r=2$ to $r=18$ gravitational radii. Broader and more
redshifted profiles correspond to smaller rings, which rotate at faster
speed and reside deeper in the gravitational well. Energy is normalized
to the unit rest energy of the line; each profile then extends from 
$g_{\rm{}min}$ to $g_{\rm{}max}$ for its corresponding parameters.
Background continuum is subtracted. Right: as on the left, but with
rings of a small (finite) radial extent of $\Delta r=1$. The rest energy
of the line is set to $6.4$~keV and a power-law continuum added to
reflect the fact that line profiles in real spectra are obtained by
considering the proper underlying continuum. Dashed lines denote the
individual components forming the prototypical spectrum; the latter is
shown by the solid line. The signature of the individual rings is
visible in the wings of the final profile. The common parameters of both
plots are: observer inclination $75^{\circ}$ (i.e.\ close to the edge-on
view), black hole spin $a=0.998$ (prograde rotation).
\label{fig1}}
\end{figure*}

\section{Relativistic line as superposition of ring profiles}
\label{sec:broadline}
Relativistic spectral lines have been modeled via various approaches,
analytical and numerical ones. But one may also ask a simpler question
about the extremal energy shifts, which basically give only the line
width and the position of the two horns rather than a detailed spectral 
shape. A double-horn spectral profile is a specific feature of a ring
positioned at a given radius. Although the calculation providing just
the extremal shifts should seemingly be easy, it has not yet been
brought out in a systematic manner. 

Fine substructures of the relativistic line from the accretion disk can
be used to constrain the inclination angle, radial emissivity 
distribution in the disk plane, and even the angular momentum of the
central black hole \citep{beckwith04}. Figure~\ref{fig1} illustrates
this by showing the formation of a model spectrum  originating as a
superposition of several ring profiles. Blue and red horns of the
separate rings rise up above the central body of the line. They can be
recognized in the wings of the total profile (relative normalization of
the rings fluxes has been set proportional to $r^{-3}$). Sharp peaks of
the spectral profile  from a narrow ring occur at the maximum and
minimum  values of the observed energy.

Let us note that the idea of studying the signatures of black holes via 
spectroscopy of radiation sources in relativistic orbital motion has a
long history \citep{cunningham73}. Theoretical light curves and spectral
line profiles were calculated including various effects of general
relativity: the frame-dragging, extreme light-bending, and multiple
images \citep[][and further references cited
therein]{kojima91,laor91,karas92,matt93,fanton97}.

We concentrate on direct evaluation of the extremal shifts, while the
available numerical tools can be used to test our results. To this end,
we employ the \textsc{ky} suit of codes \citep{dovciak04}, which
includes \textsc{kyrline} routine for the desired observed shape of a 
relativistic line. In order to achieve high accuracy of simulated lines
from very narrow annuli ($\Delta r\lesssim0.1$), we found the
\textsc{kyrline} code to be superior in the sense that the resulting
profiles do not contain artificial numerical oscillations.\footnote{The
code struggles with numerical issues only for infinitesimal rings and
very large (edge-on) inclinations, $\Delta r\lesssim0.01$ and
$i\gtrsim89^{\rm{}o}$. These limitations do not pose any problem for
objects seen at moderate and even rather high inclinations ($i\lesssim
85^{\rm{}o}$), and the new version of the code (Dov\v{c}iak et al., in
preparation) improves also these extreme cases.} This high accuracy is
important for understanding the interplay of general relativistic
effects (energy shifts and the light bending) that form the spectral
profile, because the final shape is more complicated than a simple
special-relativistic double-horn line.

\section{Extremal energy shifts from a ring}
\label{sec:extremalshifts}
\subsection{Light rays as null geodesics}
\label{sec:laightrays}
We consider a  spectral line originating from the surface of a
geometrically  thin, optically thick (standard) accretion disk
\citep[e.g.,][]{frank02}.  Propagation of photons from the disk is
treated in the limit of geometrical  optics in Kerr metric
\citep{mtw73}.\footnote{Hereafter, we express  lengths in units of
gravitational radius, 
$r_{\rm{}g}\,\equiv\,GM/c^2\,\doteq\,1.48\times10^{12}M_7~\mbox{cm}$,
where $M_7$ is the mass of the black hole in units of $10^7$ solar
masses. We use Boyer-Lindquist spheroidal coordinates, 
$(t,r,\theta,\phi)$.}

Geodesic motion is determined by three constants of motion: the total
energy $\cal E$, the azimuthal component of angular momentum $L_{\rm
z}$, and Carter's constant $Q$. For photons, null geodesics are
relevant, and for them the number of free constants can be further
reduced by re-normalizing $L_{z}$ and $Q$ with respect to energy:
$\lambda = L_{z}/{\cal E}$, $q^{2} = Q/{\cal E}^{2}$. For photons
propagating from the accretion disk towards a distant observer, the
initial point is set at a given radius in the equatorial plane of the
black hole, whereas the final point is at radial infinity, along the
view angle of the observer.

Carter's equations for light rays can be written in the integral form 
\citep{carter68},
\begin{equation}\label{carter2}
\int_{r}\frac{{\rm d}r}{\sqrt{R(r,\lambda,q^2)}} = \pm\int_{\mu}\frac{{\rm d}\mu}{\sqrt{\Theta(\mu,\lambda,q^{2})}},
\label{carter1}
\end{equation}
where
\begin{eqnarray}
    R(r,\lambda,q^2) &   =   & r^{4} + (a^{2} - \lambda^{2} - q^{2})r^{2}  \nonumber \\
&& + 2[q^{2} + (\lambda - a)^{2}]r 
 - a^{2}q^{2},\\
    \Theta(\mu,\lambda,q^{2}) &   =   & q^{2} + (a^{2} - \lambda^{2} - q^{2})\mu^{2} - a^{2}\mu^4 ,
\end{eqnarray}
$\mu=\cos\theta$, and  $a$ is the dimensionless spin of the black hole
($0\leq{a}\leq1$). The left-hand side of eq.\ (\ref{carter1}) describes
the motion in radial coordinate, while the right-hand side concerns the
latitudinal motion. These equations can be integrated in terms of
elliptic integrals \citep[e.g.,][]{rauch94,cadez98}. Roots of
polynomials $R(r)$ and $\Theta(\mu)$ correspond to turning points in the
radial and latitudinal directions, respectively. 

The radial polynomial can be expressed in the form $R = (r - r_{1})(r -
r_{2})(r - r_{3})(r - r_{4})$, where
\begin{equation}
r_{1,2} = \textstyle{\frac{1}{2}}F \pm \textstyle{\frac{1}{2}}D_{-}^{1/2},
\quad
r_{3,4} = -\textstyle{\frac{1}{2}}F \pm \textstyle{\frac{1}{2}}D_{+}^{1/2}
\end{equation}
are roots of the polynomial $R(r)$.
The latitudinal polynomial adopts the form 
$\Theta(\mu) = a^{2}(\mu^{2}_{-} + \mu^{2})(\mu^{2}_{+} - \mu^{2})$, with
$q^{2} > 0$ and the roots
\begin{equation}
\mu^{2}_{\pm} = \frac{1}{2a^{2}}\left[\left(G^{2} + 4a^{2}q^{2}\right)^{1/2} \mp G\right].
\end{equation}
We denoted constants:
\begin{eqnarray*}
A   &   \equiv   &   (a^{2} - \lambda^{2} - q^{2}),\\
B   &   \equiv   &   (a - \lambda)^{2} + q^{2},\\
C   &   \equiv   &   A^{2} - 12a^{2}q^{2},\\
D   &   \equiv   &   2A^{3} + 72a^{2}q^{2}A + 108B^{2},\\
E   &   \equiv   &   \textstyle{\frac{1}{3}}\,[(\textstyle{\frac{1}{2}}E_+)^{1/3}+(\textstyle{\frac{1}{2}}E_-)^{1/3}],\\
F   &   \equiv   &   (E-\textstyle{\frac{2}{3}}A)^{1/2},\\
G   &   \equiv   &   \lambda^{2} + q^{2} - a^{2},
\end{eqnarray*}
with 
$D_{\pm} = -\frac{4}{3}A - E \pm 4BF^{-1}$,
and
$E_{\pm} = D \pm (D^{2}-4C^{3})^{1/2}$.

\subsection{Photon energy on arrival to observer}
The energy shift is defined as ratio of observed $E_{\rm o}$ to
emitted $E_{\rm e}$ photon energy
\begin{equation}
g = \frac{E_{\rm o}}{E_{\rm e}}.
\label{eq:g}
\end{equation}
The emission source orbits with four-velocity
$\bmath{u} = u^{t}(1,0,0,\Omega)$, where
\begin{equation}
u^{t} = \left[1 - 2r_{\rm e}^{-1}\left(1 - a\Omega\right)^{2} - \left(r^{2}_{\rm e} + a^{2}\right)\Omega^{2}\right]^{-1/2},
\end{equation}
with $\Omega(r_{\rm e}) =(r^{3/2}_{e} + a)^{-1}$. Introducing the
angular velocity into eq.\ (\ref{eq:g}), the energy shift is
\begin{equation}
\label{fce_g}
g = \frac{1}{u^{t}}\frac{1}{1 - \lambda\Omega}.
\end{equation}
We look for extremal values of the function (\ref{fce_g}). The
definition domain of the energy shift as a function of specific angular
momentum, $g(\lambda)$, is an interval $(\lambda_{\rm min},\lambda_{\rm
max})$, which is constrained by the condition of photon reaching the
observer. 

Even though the equations of the previous section are fairly well known
and were discussed in various papers, the extremes of the redshift
function $g$ are not so easy to write in an analytical way. Let us
remark that an elegant way of determining the energy shifts was derived
by \citet{schee05} in terms of the light emission loss cone. However,
their approach allows only to find the extremal energy shifts of {\em
all\/} photons emitted from the source at a given position. This
includes also those which follow indirect light rays and cross the disk
plane. Although the family of direct light rays have generally a simpler
shape than the indirect rays, the additional condition prevents us from
using the loss cone method to determine the range of energy shifts for a
source in the accretion disk.

Only certain combinations of roots and turning points are relevant for
light rays involved in our discussion, i.e.\ those starting from the
equatorial ring and reaching a given observer, not crossing the
equatorial plane for the second time (we assume that the rays crossing
the equatorial plane are obstructed by the disk). We obtain the
following combinations.

\subsubsection*{The radial integral, four real roots.}
We find
\begin{equation}
\int\limits^{\infty}_{r_{\rm e}}\frac{{\rm d}r}{\sqrt{R(r)}} = g_{r}\big[F(\varphi_{\rm o},k_{r}) \pm F(\varphi_{\rm e},k_{r})\big],
\label{eq:fourreal}
\end{equation}
where $F(\varphi,k_{r})$ is the elliptical integral of the first kind
\citep{byrd71},
\begin{eqnarray}
   g_{r}(\lambda,q^{2}) &  =  & 2(r_{1} - r_{3})^{-1/2}(r_{2} - r_{4})^{-1/2} ,\nonumber\\
   k_{r}(\lambda, q^{2}) &  =  & \frac{(r_{2} - r_{3})(r_{1} - r_{4})}{(r_{1} - r_{3})(r_{2} - r_{4})}\,,\nonumber\\
   \varphi_{\rm o}(\lambda,q^{2}) &  =  & \arcsin\left(\frac{r_{2}- r_{4}}{r_{1}  - r_{4}}\right)^{1/2} ,\nonumber\\
   \varphi_{\rm e}(\lambda,q^{2}) &  =  & \arcsin\left[\frac{(r_{2}-r_{4})(r_{\rm e} - r_{1})}{(r_{1}  - r_{4})(r_{\rm e} - r_{2})}\right]^{1/2}      .\nonumber
\end{eqnarray}
The upper sign in eq.~(\ref{eq:fourreal}) refers to the case of light
rays passing through a turning point in the radial direction; the
lower sign refers to those with no radial turning point.

\subsubsection*{The radial integral, two complex roots.} 
We set $r_{1,2} = u \pm {\rm i}v$ to denote the complex roots, and
$r_{3,4}$ to be the real roots. Then, $u = \frac{1}{2}F$, $v =
\frac{1}{2}{|D_{-}|^{1/2}}$. The radial integral adopts the form
\begin{equation}
\int\limits^{\infty}_{r_{\rm e}}\frac{{\rm d}r}{\sqrt{R(r)}} = g_{r}\big[F(\varphi_{\rm o},k_{r}) - F(\varphi_{\rm e},k_{r})\big],
\end{equation}
where
\begin{eqnarray*}
g_{r}(\lambda,q^{2}) &  =  &  A_{c}^{-1/2}B_{c}^{-1/2},\\
k_{r}(\lambda, q^{2}) &  =  &  \frac{(A_{c} + B_{c})^{2} - (r_{3} - r_{4})^{2}}{4A_{c}B_{c}},\\
\varphi_{\rm o}(\lambda,q^{2}) &  =  &  \arccos\left[\frac{A_{c} - B_{c}}{A_{c} + B_{c}}\right],\\
\varphi_{\rm e}(\lambda,q^{2}) &  =  &  \arccos\left[\frac{(A_{c} - B_{c})r_{\rm e} + r_{3}B_{c} - r_{4}A_{c}}{(A_{c} + B_{c})r_{\rm e} - r_{3}B_{c} - r_{4}A_{c}}\right] ,\\
A_{c}(\lambda,q^{2}) &  =  &  \left[(r_{3} - u)^{2} + v^{2}\right]^{1/2},\\
B_{c}(\lambda,q^{2}) &  =  &  \left[(r_{4} - u)^{2} + v^{2}\right]^{1/2}.
\end{eqnarray*}

\subsubsection*{The latitudinal integral.} 
The latitudinal integral can be written in the form
\begin{equation}
\int\limits^{\mu_{\rm e}}_{0}\frac{{\rm d}\mu}{\sqrt{\Theta(\mu,\lambda,q^{2})}} = \frac{g_{\mu}}{a}F(\psi,k_{\mu}),
\end{equation}
assuming that the light ray has no latitudinal turning point.
Otherwise, the appropriate form of the integral is
\begin{equation}
\int\limits^{\mu_{\rm e}}_{0}\frac{{\rm d}\mu}{\sqrt{\Theta(\mu,\lambda,q^{2})}} = \frac{g_{\mu}}{a}[2K(k_{\mu}) - F(\psi,k_{\mu})],
\end{equation}
where
\begin{eqnarray*}
g_{\mu}(\lambda,q^{2}) &  =  & (\mu^{2}_{+} + \mu^{2}_{-})^{-1/2},\\
k_{\mu}(\lambda,q^{2}) &  =  & \mu^{2}_{+}\,(\mu^{2}_{+} + \mu^{2}_{-})^{-1},\\
\psi(\lambda,q^{2}) &  =  & \arcsin\left[\frac{\mu^{2}_{\rm o}(\mu^{2}_{+} + \mu^{2}_{-})}{\mu^{2}_{+}(\mu^{2}_{\rm o} + \mu^{2}_{-})}\right]^{1/2},
\end{eqnarray*}
and $K(k_{\mu}) = F(\frac{\pi}{2},k_{\mu})$.

\begin{figure*}[tbh!]
\includegraphics[angle=-90,width=.49\textwidth]{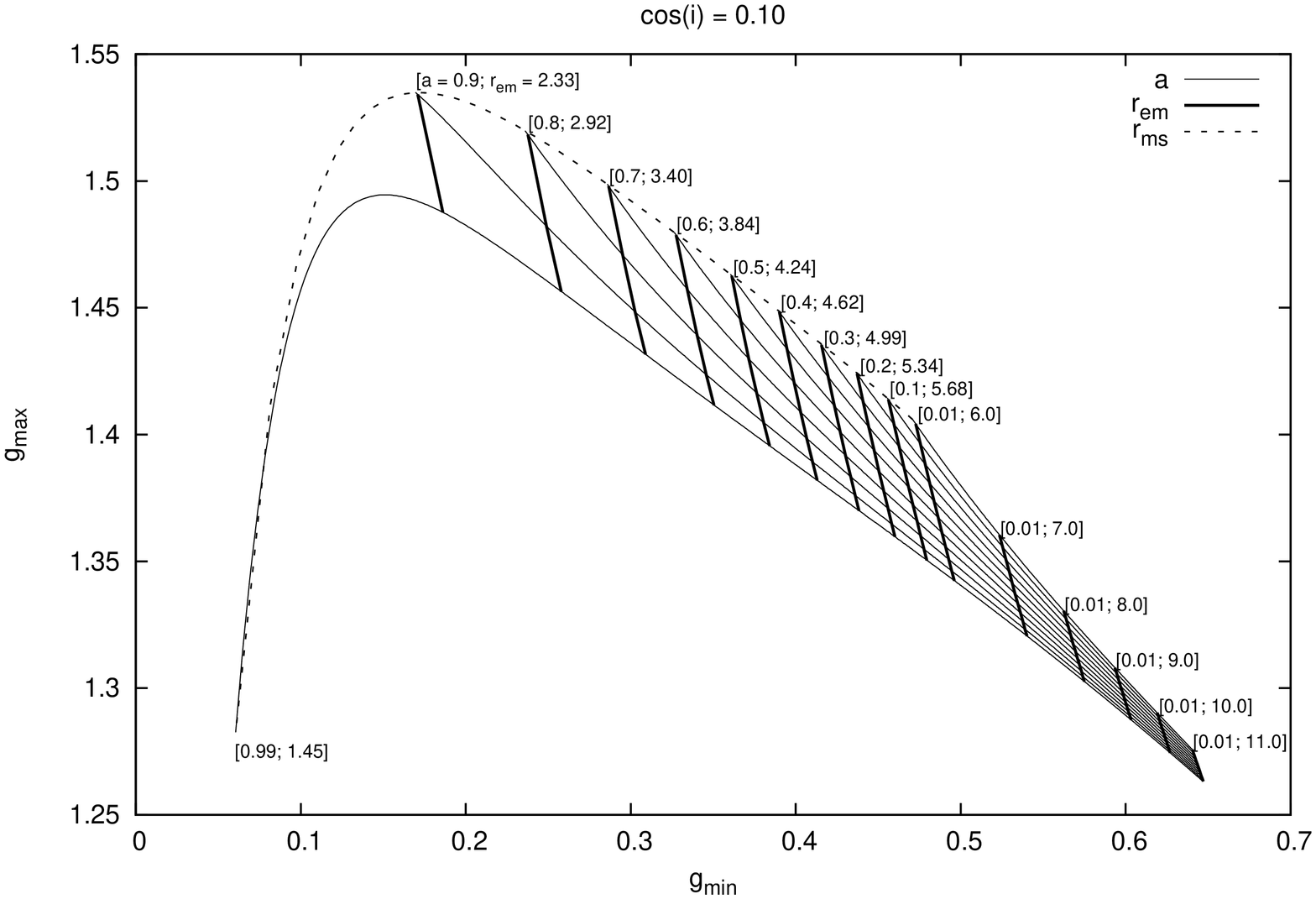}
\hfill
\includegraphics[angle=-90,width=.49\textwidth]{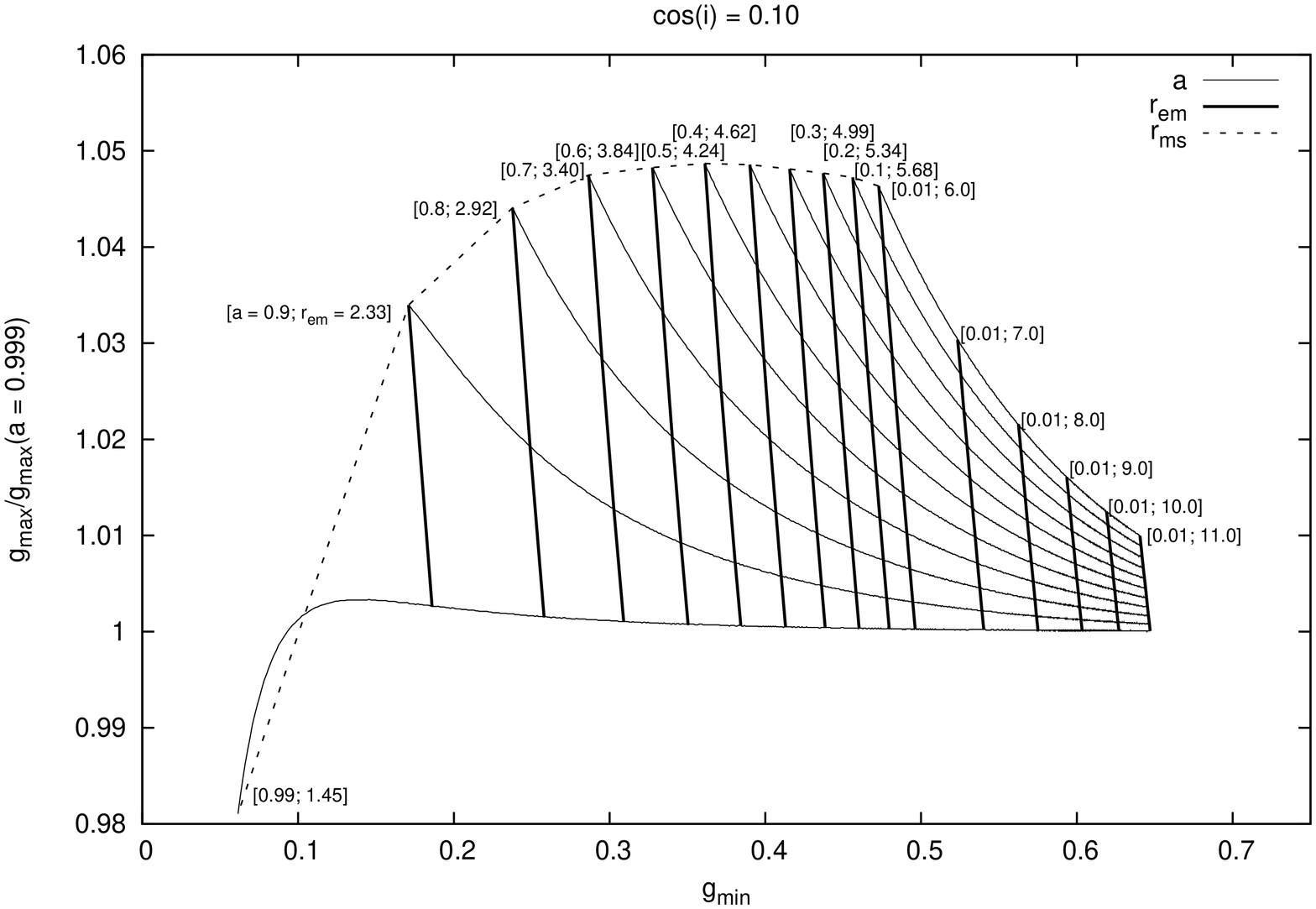}
\includegraphics[angle=-90,width=.49\textwidth]{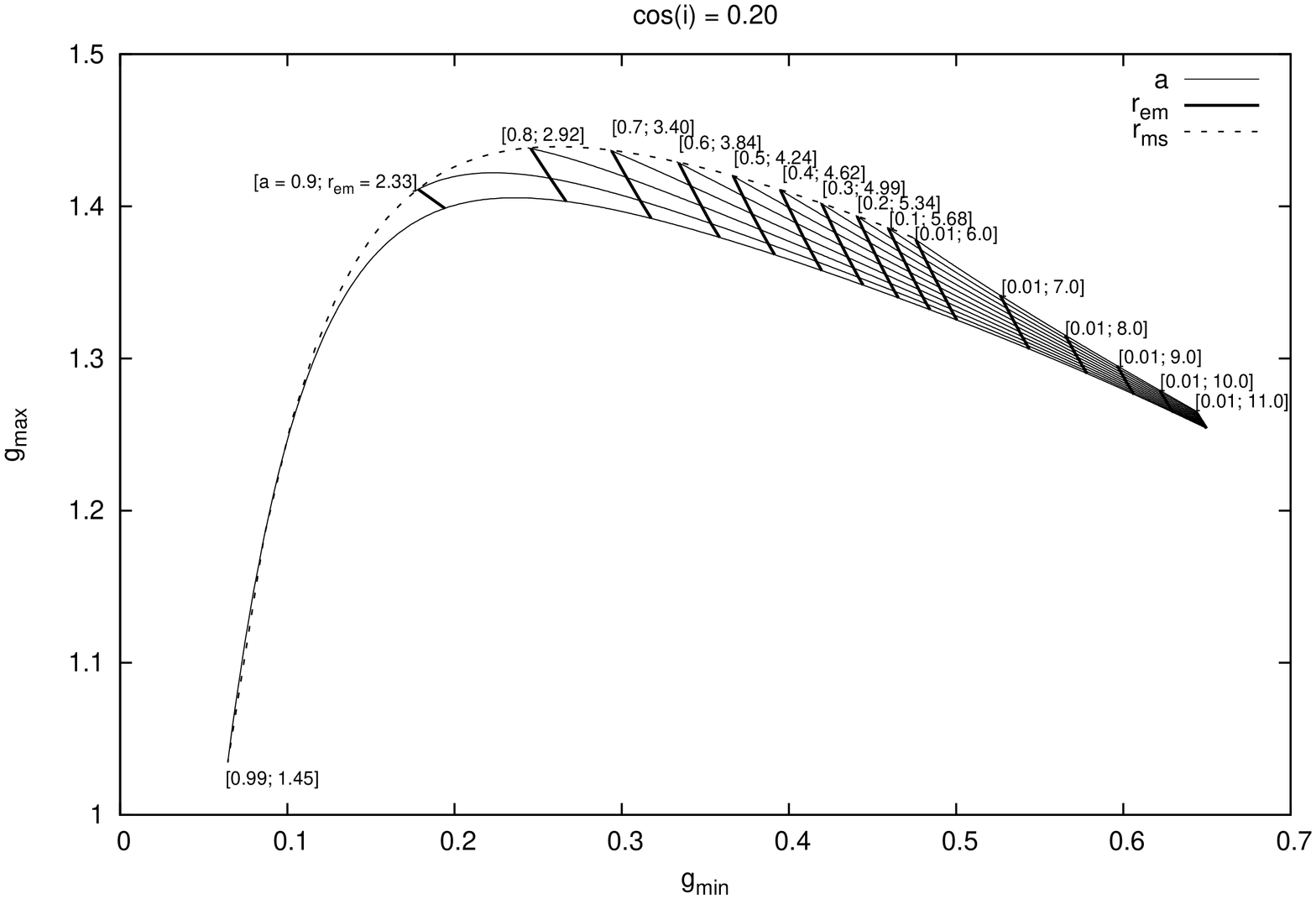}
\hfill
\includegraphics[angle=-90,width=.49\textwidth]{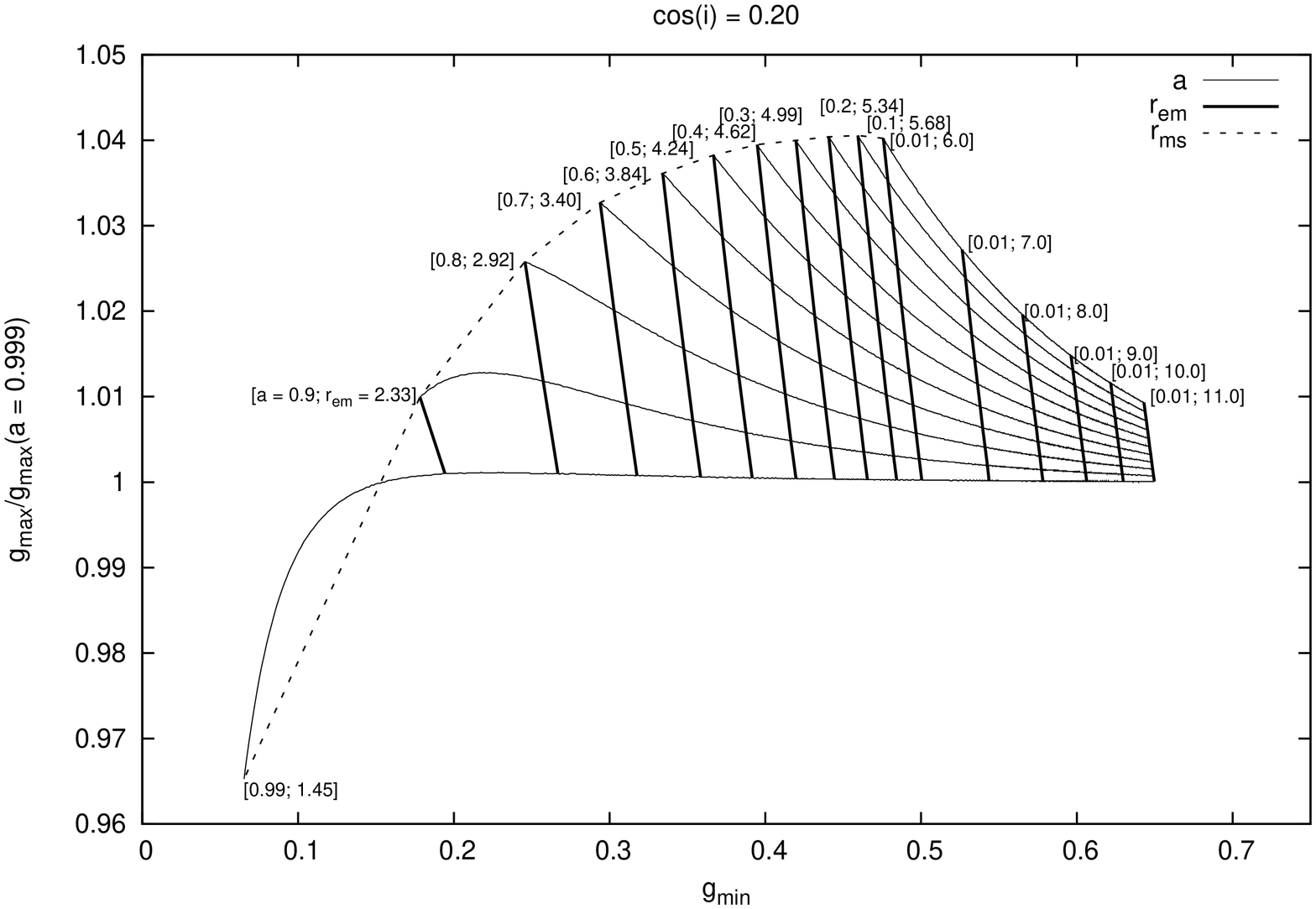}
\caption{Extremal shifts of the 
observed photon energy for different view angles of the observer ($\cos
i$, given on top of each panel). Left panels: The magnitude of
$g_{\rm max}$ versus $g_{\rm min}$. Right panels: As on the left but
showing the normalized values on the ordinate (for better clarity of the
plot, especially at lower inclinations). Each pair of $g_{\rm{}max}$, 
$g_{\rm{}min}$ values gives the corresponding emission radius $r_{\rm{}em}$
and the black hole spin $a$. Curves of constant $r_{\rm{}em}$ and the spin 
$a$ are distinguished by different line width (the values are written in
brackets). ISCO radius $r=r_{\rm{}ms}$ defines one boundary of the plot
(dotted curve). See the text for details.
\label{fig2}}
\end{figure*}

\begin{figure*}[tbh!]
\includegraphics[angle=-90,width=.49\textwidth]{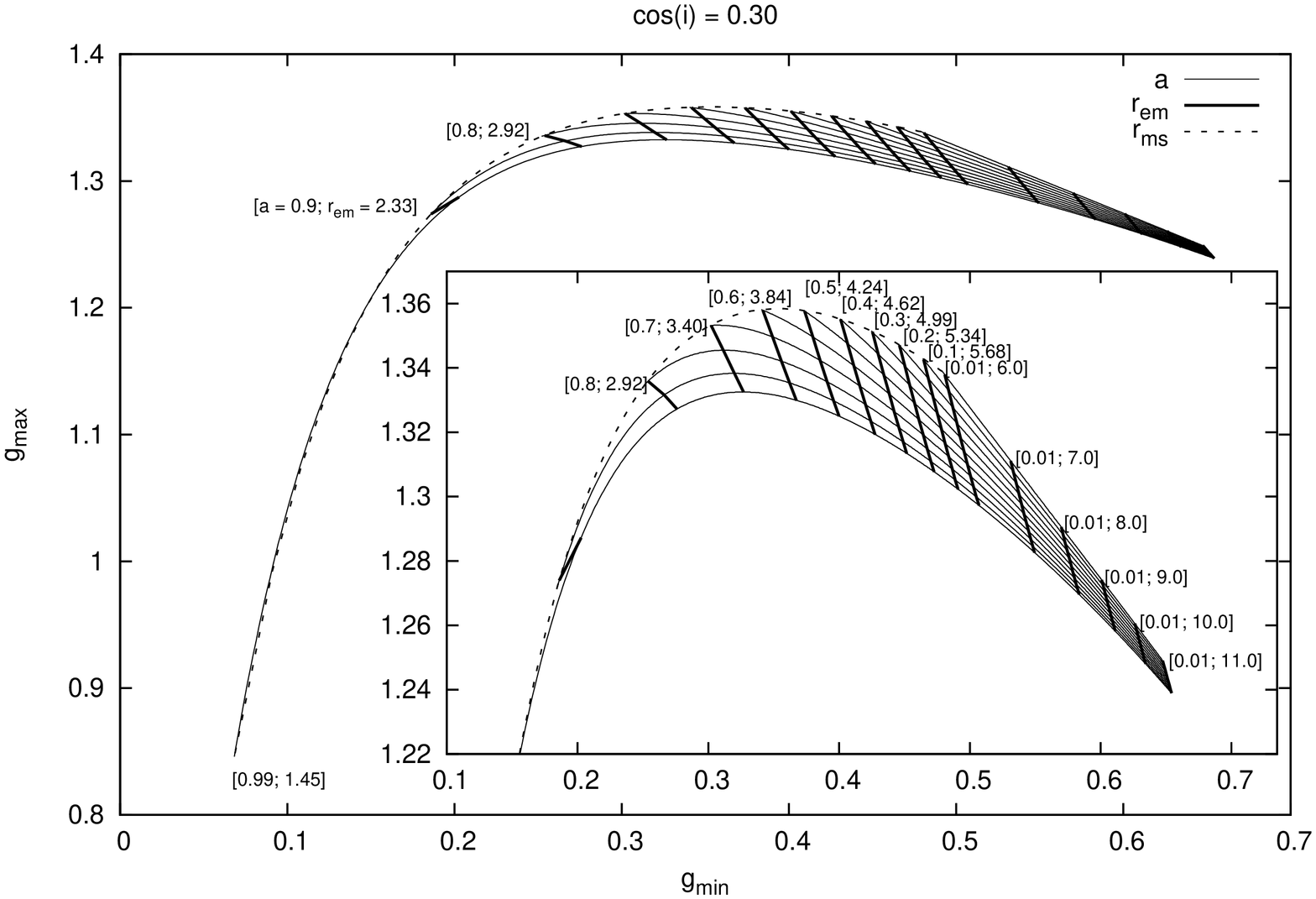}
\hfill
\includegraphics[angle=-90,width=.49\textwidth]{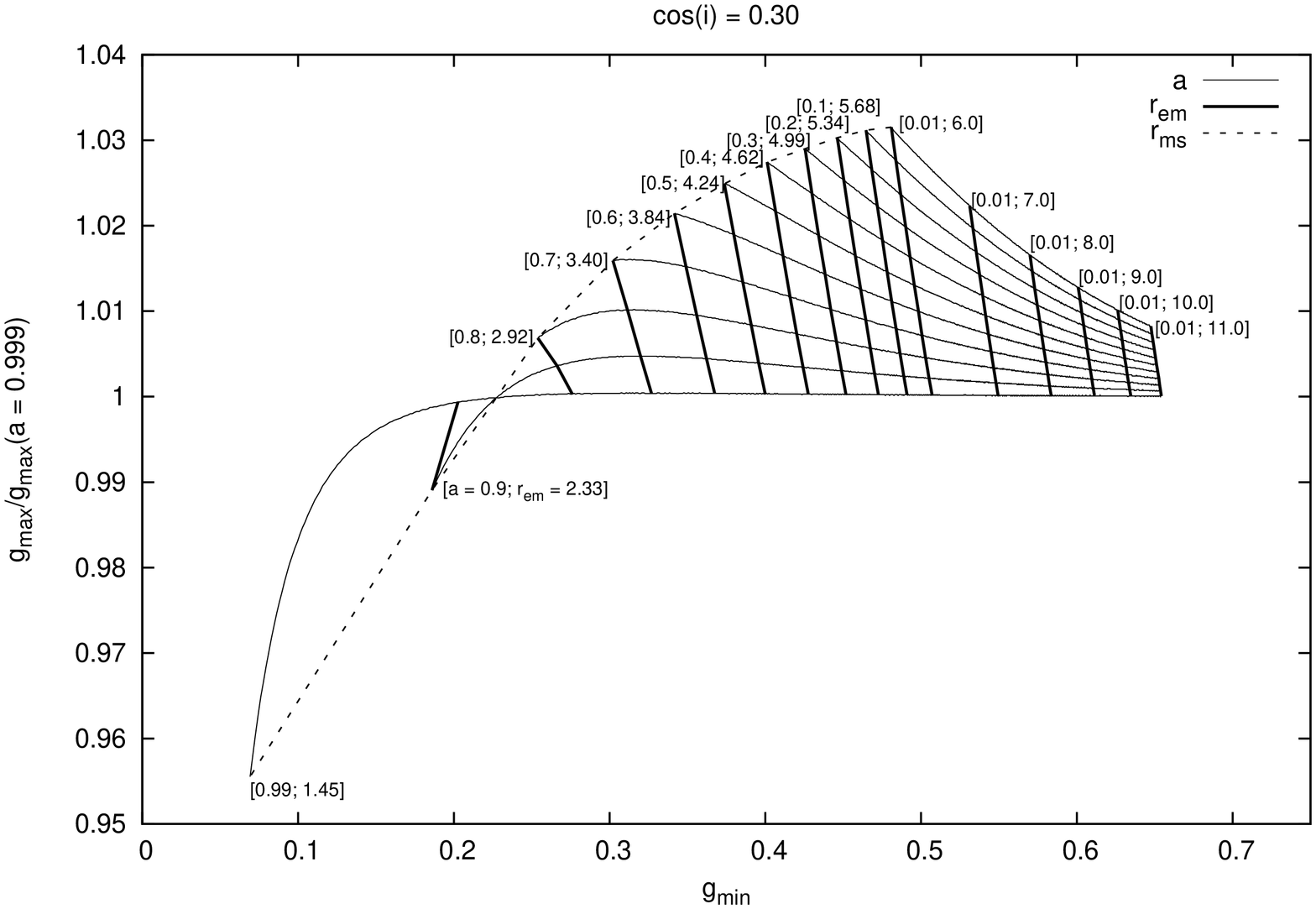}
\includegraphics[angle=-90,width=.49\textwidth]{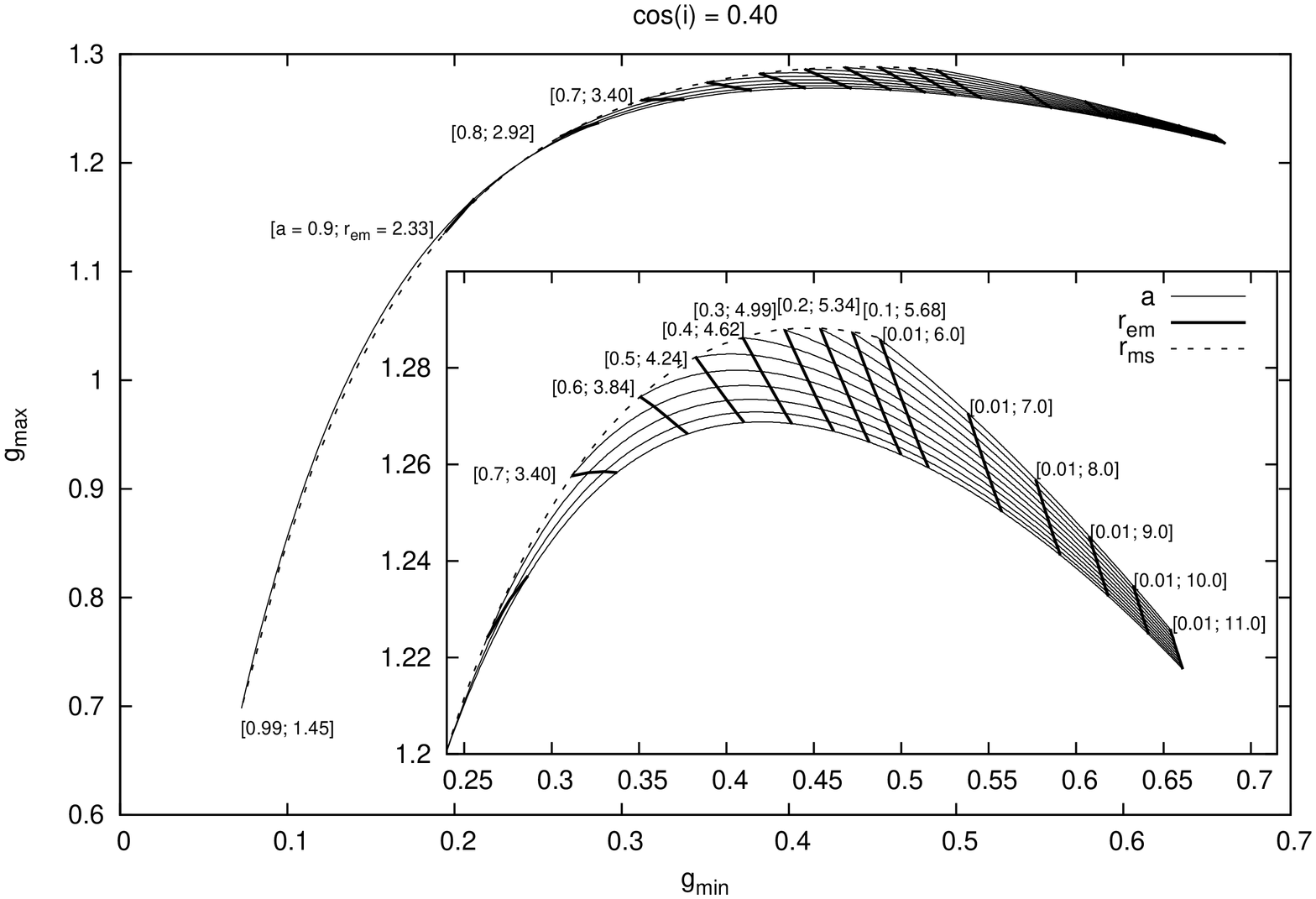}
\hfill
\includegraphics[angle=-90,width=.49\textwidth]{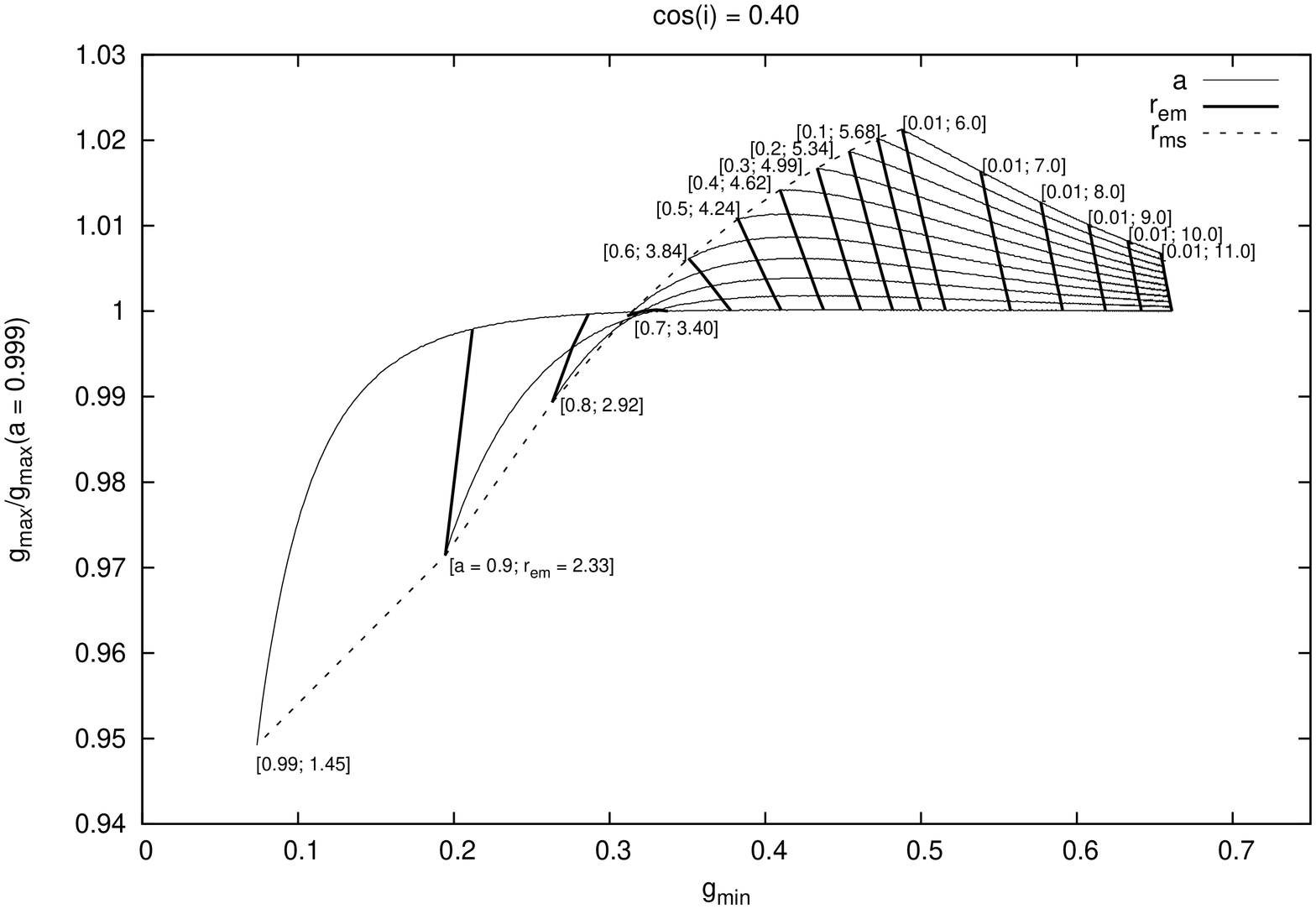}
\caption{As in the previous figure, but for lower inclination
angles. By decreasing the inclination the dependency on the black hole
spin becomes less prominent, and so the curves for different $a$ get
closer to each other. Therefore, in the left column we show an inset
where the relevant part of the plot is enlarged.
\label{fig3}}
\end{figure*}

\begin{figure*}[tbh!]
\includegraphics[angle=-90,width=.49\textwidth]{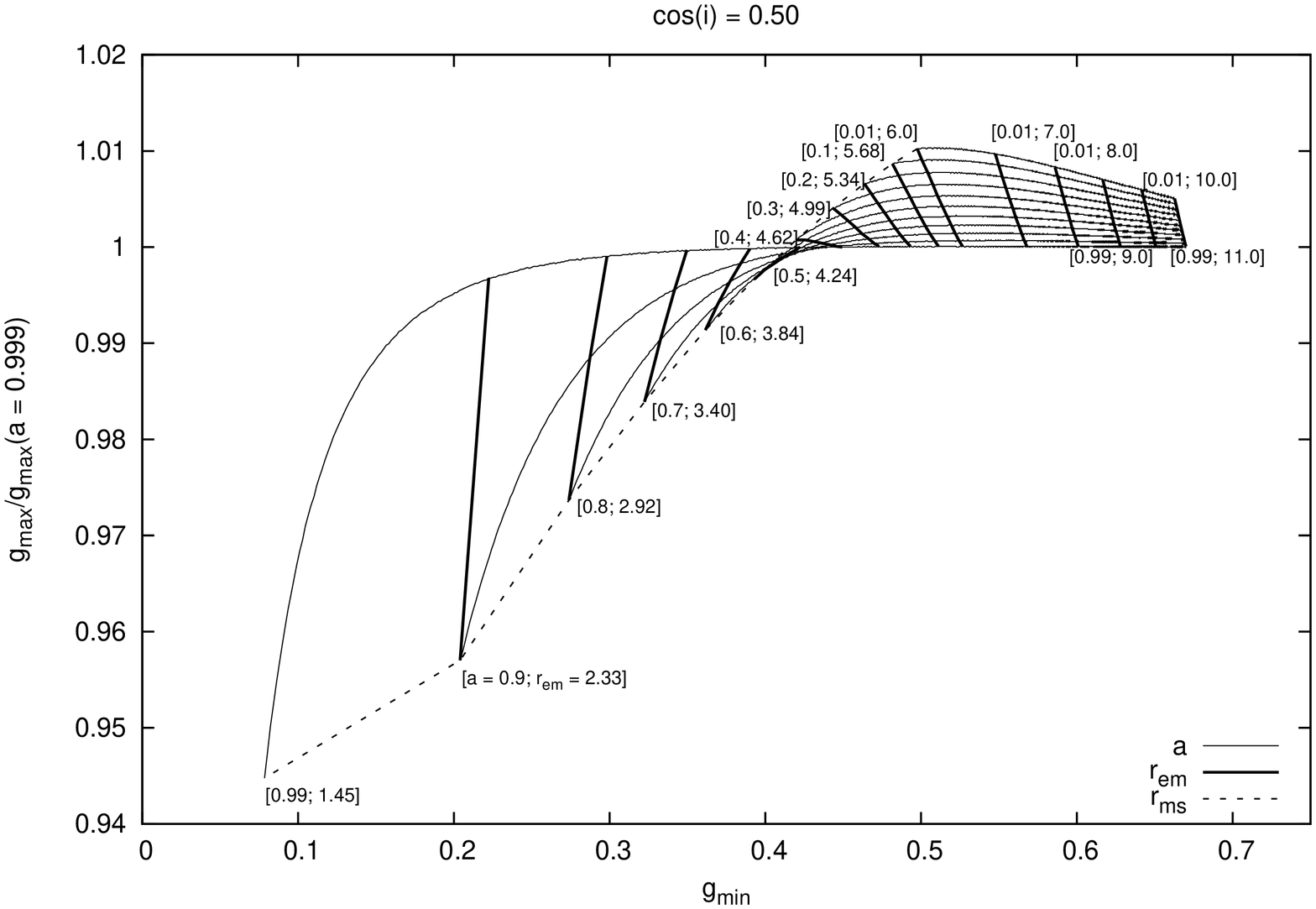}
\hfill
\includegraphics[angle=-90,width=.49\textwidth]{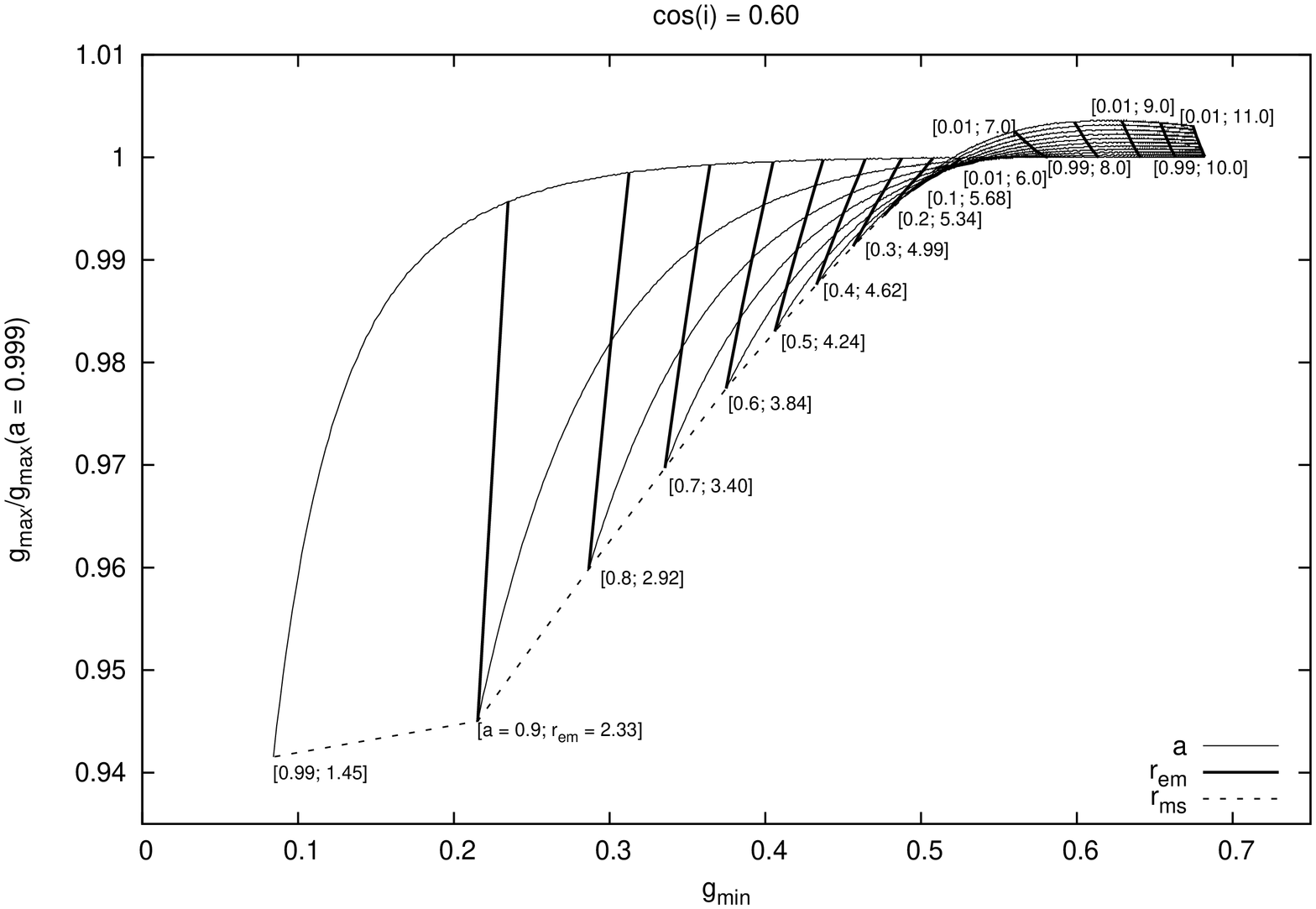}
\includegraphics[angle=-90,width=.49\textwidth]{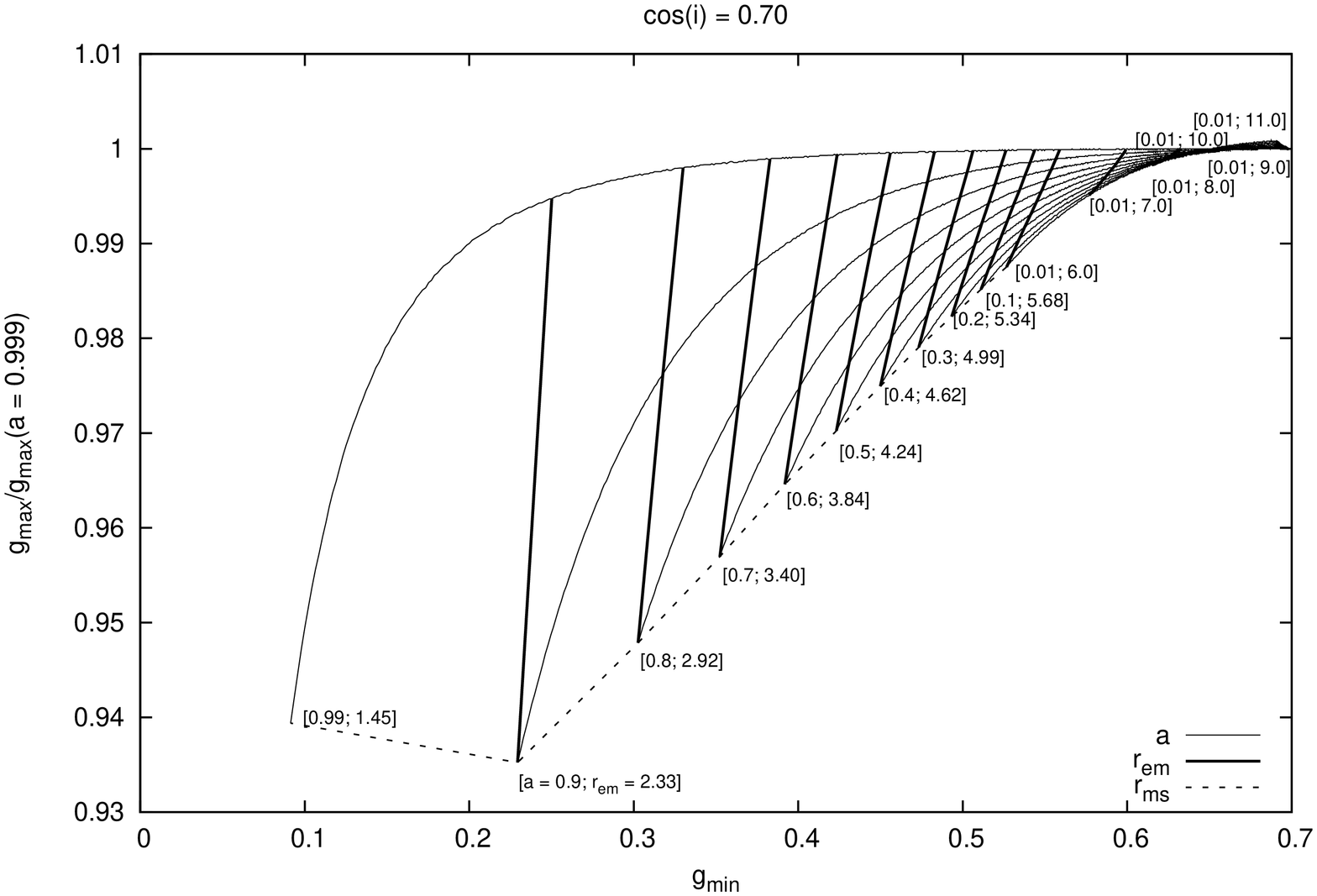}
\hfill
\includegraphics[angle=-90,width=.49\textwidth]{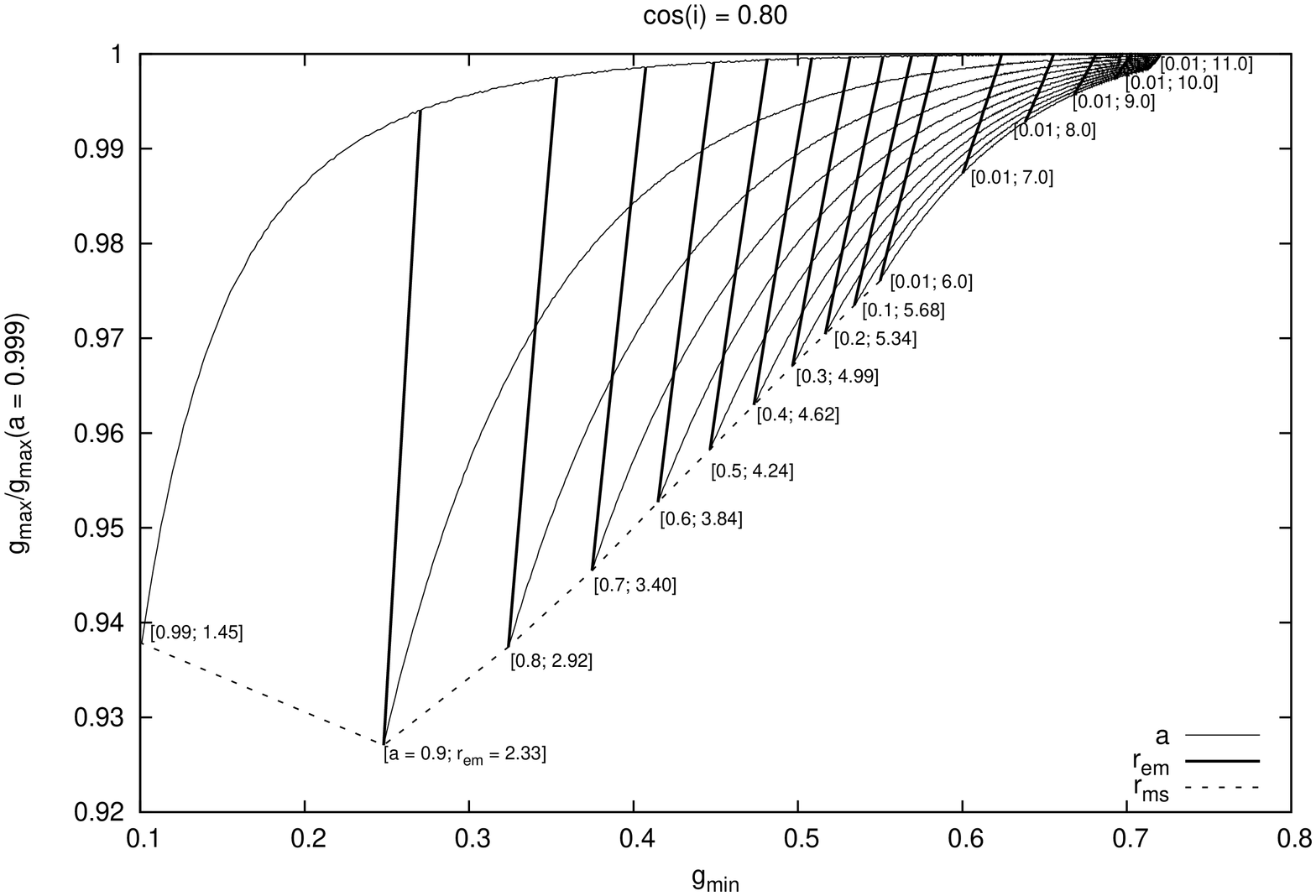}
\includegraphics[angle=-90,width=.49\textwidth]{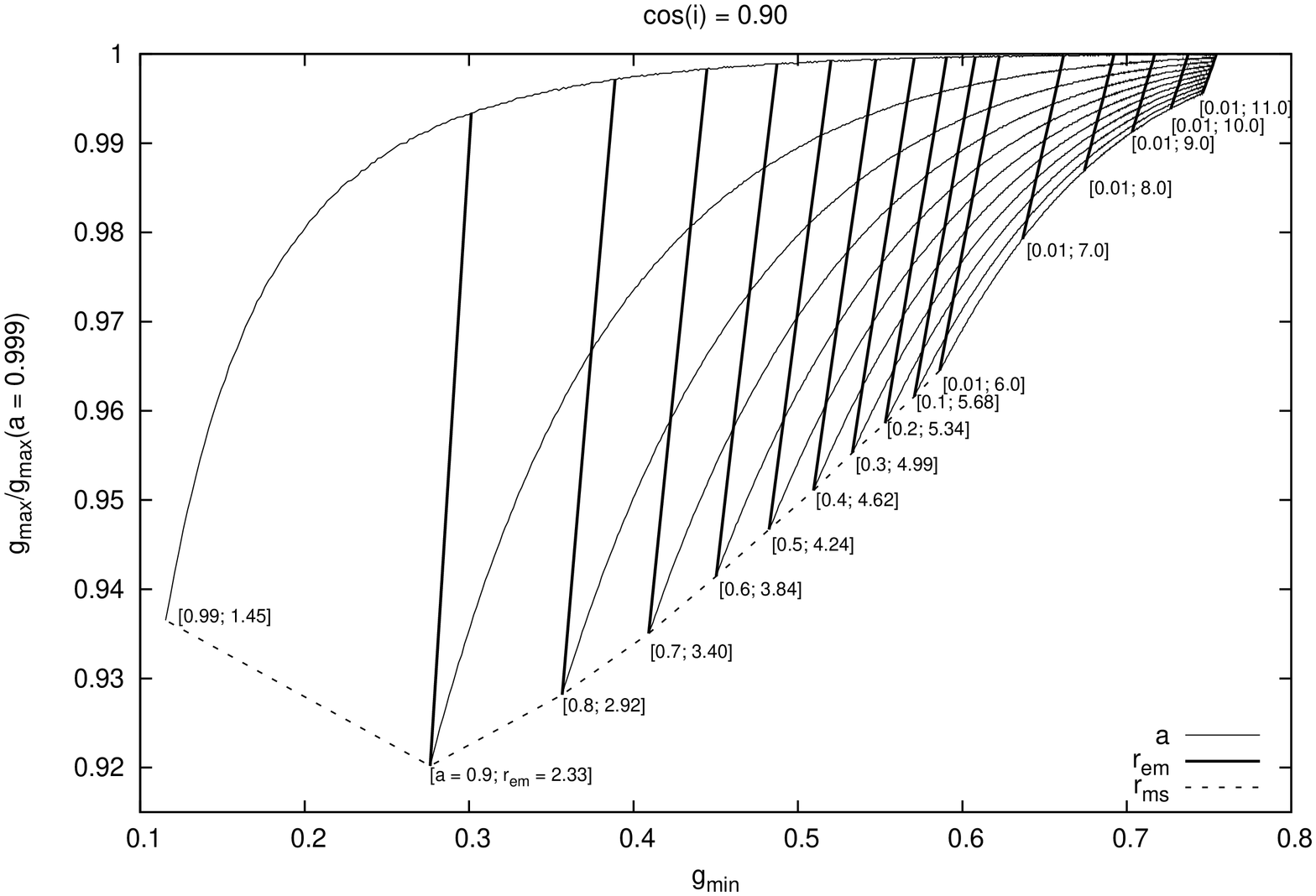}
\hfill
\includegraphics[angle=-90,width=.49\textwidth]{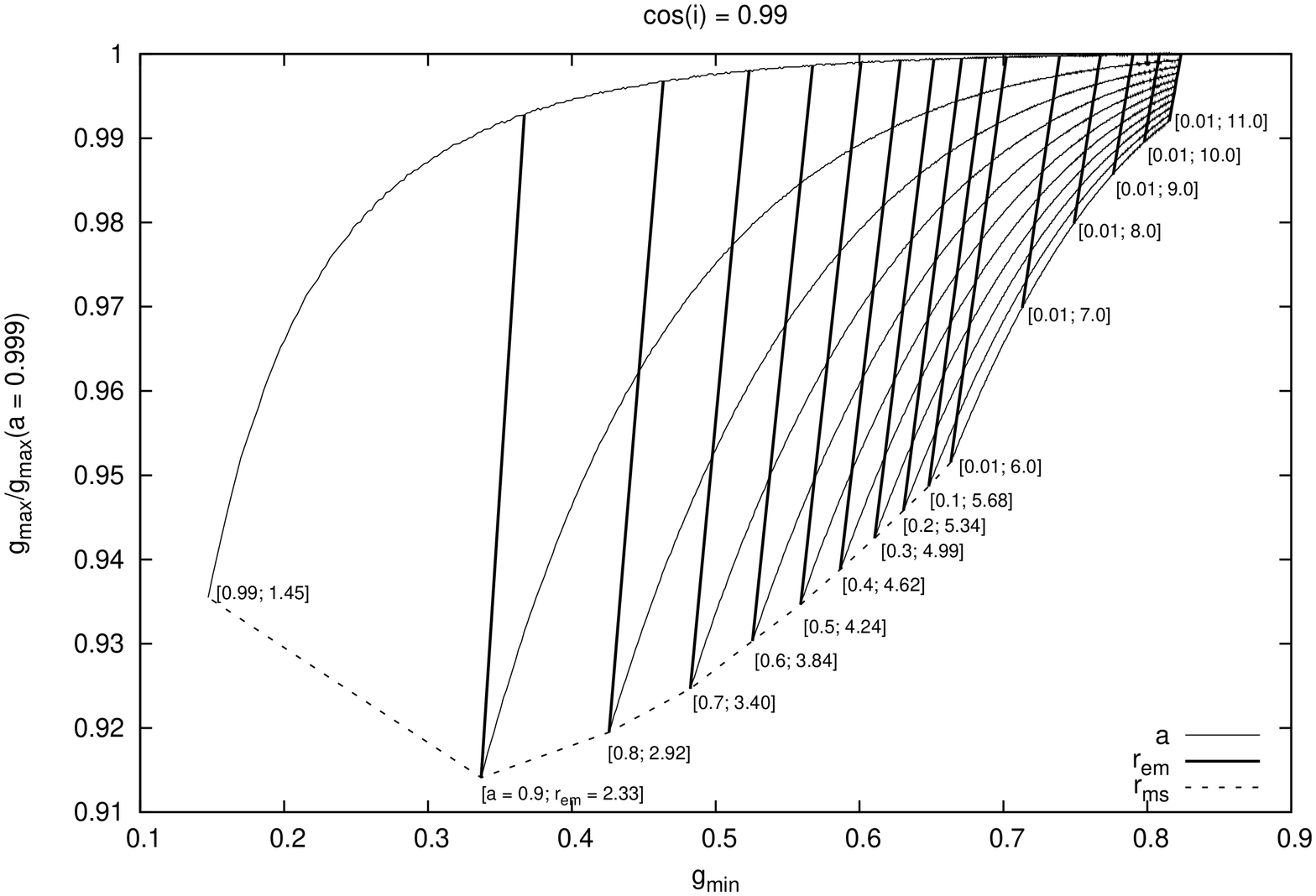}
\caption{Normalized 
graphs of $g_{\rm{}max}$ vs.\ $g_{\rm{}min}$ for low inclinations, i.e.\
close to the face-on view of the disk plane. Bottom-right panel
corresponds to the view of the disk along the rotation axis.
\label{fig4}}
\end{figure*}

\begin{figure*}[tbh!]
\begin{center}
\includegraphics[angle=-90,width=0.75\textwidth]{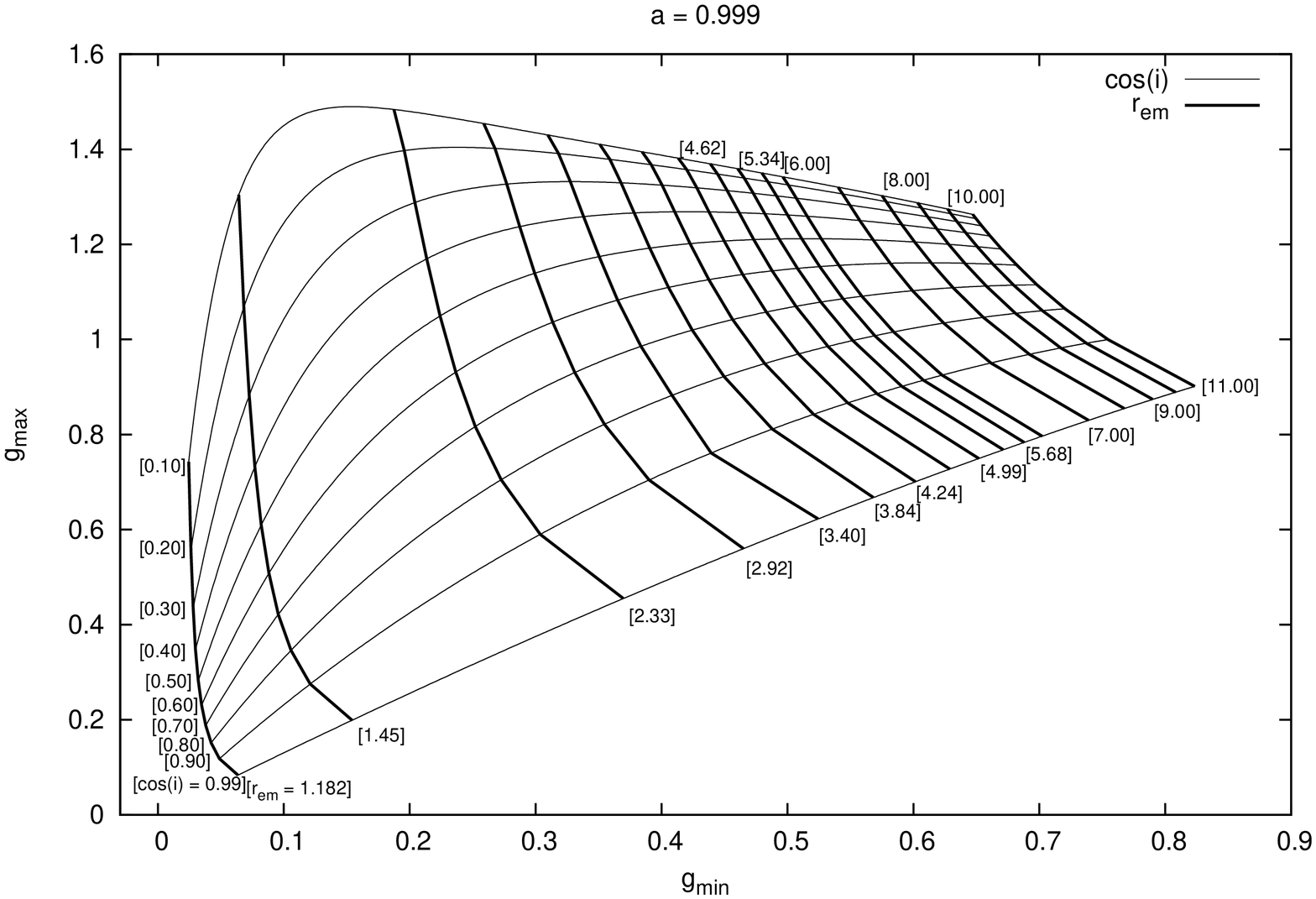}
\end{center}
\caption{Extremal energy shift for $a=0.999$. Set of contours of
constant inclination angle and of constant emission radius are shown.
This plot gives the scaling factor $g_{\rm{}max}(a=0.999)$ of the
normalized graphs in Figures \ref{fig2}--\ref{fig4}.
\label{fig5}}
\end{figure*}

\subsection{Iterative solution for extremal shifts}
\label{sec:iterative}
We search for the extremal values $g_{\rm min}$, $g_{\rm max}$ of the
redshift function (\ref{fce_g}), under a simultaneous constraint by eq.\
(\ref{carter2}). Lagrange multipliers provide a suitable strategy for 
finding the constrained extremal values. To this end, the multipliers
$\alpha$ are defined by the relation
\begin{eqnarray}
\Lambda(\lambda,q^{2},\alpha) &   =   & \frac{1}{u^{t}}\frac{1}{1 - \lambda\Omega} - \alpha\int\limits^{\infty}_{r_{\rm e}}\frac{{\rm d}r}{\sqrt{R(r,\lambda,q^{2})}} \nonumber\\
&&\,+ \alpha\int\limits^{\mu_{\rm o}}_{0}\frac{{\rm d}\mu}{\sqrt{\Theta(\mu,\lambda,q^{2})}}\,,
\end{eqnarray}
where partial derivatives of the Lagrange function
$\Lambda(\lambda,q^{2},\alpha)$ with respect to $\lambda$, $q^2$ and
$\alpha$ must vanish identically. The latter condition yields two
coupled equations for the unknowns $\lambda$ and $q^2$,
\begin{equation}
\label{eq1}
f_1 = \int\limits^{\infty}_{r_{\rm e}}\frac{{\rm d}r}{\sqrt{R(r,\lambda,q^2)}} - \int\limits^{\mu_{\rm o}}_{0}\frac{{\rm d}\mu}{\sqrt{\Theta(\mu,\lambda,q^{2})}} = 0,
\end{equation}
and
\begin{eqnarray}
\label{eq2}
f_2 &   =   & \frac{\partial f_1}{\partial q^{2}} = \frac{\partial}{\partial q^{2}}\left[\int\limits^{\infty}_{r_{\rm e}}\frac{{\rm d}r}{\sqrt{R(r,\lambda,q^2)}} 
\right. \nonumber \\ 
&&\left.-\int\limits^{\mu_{\rm o}}_{0}\frac{{\rm d}\mu}{\sqrt{\Theta(\mu,\lambda,q^{2})}}\right] = 0.
\end{eqnarray}
The value of $\lambda$ conforming to eqs.\ (\ref{eq1})--(\ref{eq2})
corresponds to the desired extremes of the energy shift.

In order to evaluate the extremal values of $g$, we solve the set 
(\ref{eq1})--(\ref{eq2}) using the Newton-Raphson method.
To this end we write Taylor expansion about the root neighborhood,
\begin{eqnarray}
f_1(\lambda,q^2) &   =   & 0 = f_1(\lambda_{\rm n},q^2_{\rm n}) + (\lambda - \lambda_{\rm n})\frac{\partial f_1}{\partial \lambda}(\lambda_{\rm n},q^2_{\rm n}) 
\nonumber\\
&&+ (q^2 - q^2_{\rm n})\frac{\partial f_1}{\partial q^2}(\lambda_{\rm n},q^2_{\rm n}) 
\nonumber\\
&&+ {\cal O}(\lambda - \lambda_{\rm n})^2 + {\cal O}(q^2 - q^2_{\rm n})^2,\\
f_2(\lambda,q^2) &   =   & 0 = f_2(\lambda_{\rm n},q^2_{\rm n}) + (\lambda - \lambda_{\rm n})\frac{\partial f_2}{\partial \lambda}(\lambda_{\rm n},q^2_{\rm n})
\nonumber\\
&& + (q^2 - q^2_{\rm n})\frac{\partial f_2}{\partial q^2}(\lambda_{\rm n},q^2_{\rm n}) 
\nonumber\\
&&+ {\cal O}(\lambda - \lambda_{\rm n})^2 + {\cal O}(q^2 - q^2_{\rm n})^2,
\end{eqnarray}
where $n$ is the order of the expansion (to be determined by the desired
accuracy of the solution). We define $\Delta \lambda_{\rm n} =
\lambda - \lambda_{\rm n}$ and $\Delta q^2_{\rm n} = q^2 - q^2_{\rm
n}$ to obtain two relations for $\Delta \lambda_{\rm n}$ and $\Delta
q^2_{\rm n}$,
\begin{equation}\label{eq3}
\Delta \lambda_{\rm n}\frac{\partial f_1}{\partial \lambda}(\lambda_{\rm n},q^2_{\rm n}) 
+ \Delta q^2_{\rm n}\frac{\partial f_1}{\partial q^2}(\lambda_{\rm n},q^2_{\rm n})\approx 
- f_1(\lambda_{\rm n},q^2_{\rm n}),
\end{equation}
\begin{equation}\label{eq4}
\Delta \lambda_{\rm n}\frac{\partial f_2}{\partial \lambda}(\lambda_{\rm n},q^2_{\rm n}) 
+ \Delta q^2_{\rm n}\frac{\partial f_2}{\partial q^2}(\lambda_{\rm n},q^2_{\rm n})\approx 
- f_2(\lambda_{\rm n},q^2_{\rm n}),
\end{equation}
where
\begin{equation}
\Delta q^2_{\rm n} = \frac{f_{1}\,f_{2,\lambda} - f_{2}\,f_{1,\lambda}}{f_{1,\lambda}\,f_{2,q^2} - f_{1,q^2}\,f_{2,\lambda}},
\end{equation}
\begin{equation}
\Delta \lambda_{\rm n} = \frac{-f_{1} - \Delta q^2_{\rm n}\,f_{1,q^2}}{f_{1,\lambda}}.
\end{equation}

Eqs.\ (\ref{eq3})--(\ref{eq4}) are linear in $\Delta \lambda_{\rm n}$
and  $\Delta q^2_{\rm n}$. The solution can be found by successive iterations,
\begin{equation}
\lambda_{\rm n + 1} = \lambda_{\rm n} + \Delta \lambda_{\rm n}, 
\quad
q^2_{\rm n + 1} = q^2_{\rm n} + \Delta q^2_{n}.
 \end{equation}

Results are plotted in Figures \ref{fig2}--\ref{fig4}, where we show
the extremal shifts as a function of the two main parameters:
(i)~dimensionless spin of the black hole ($0\leq a \leq 1$); and
(ii)~emission radius $r_{\rm em}$ of the ring  ($r_{\rm ms}\leq r_{\rm
em}$, expressed in units of gravitational radii). The inclination angle
$i$ stands as a third parameter, which we keep fixed in each of the
figures (see the cosine of inclination on top of each panel, $0\leq i
\leq 90^{\rm o}$; edge-on view of the disk corresponds to $i=90^{\rm
o}$).

The behavior of the curves is determined by the interplay of Doppler 
effect, strong-gravity lensing, and light aberration near the black hole. 
By increasing $r$ (with $a$ and $i$ fixed) both $g_{\rm{}min}$ and
$g_{\rm{}max}$ increase when $i$ is small, showing that the dominant factor 
is the gravitational redshift rather than the relativistic beaming. On the
other hand, the latter becomes important for large inclinations.

The method of solution is efficient enough and it allows us to explore
parameters in a systematic way. On the other hand, because the parameter
space is quite rich and the plots contain wealth of information, one may
need to get accustomed to the actual meaning of the presented curves.
Broadly speaking, the approaching side of the ring produces photons
around $g_{\rm{}max}$ energy, whereas the receding part gives
$g_{\rm{}min}$ for the given radius and spin. These trends are further
influenced by the overall gravitational redshift, which eventually
prevails as the emission radius approaches the horizon, and the light
bending effect, which enhances the signal from a region of the disk
around the radiation caustic at high view angles.

We remind the reader, that the extremal shifts $g_{\rm{}min}$ and
$g_{\rm{}max}$ play a {\em role of observable quantities}. It is 
convenient to have them given directly on the axes. Given
$g_{\rm{}min}$, $g_{\rm{}max}$ one can immediately find the
corresponding values of the emission radius and the black hole spin. The
set of Figures \ref{fig2}--\ref{fig4} covers the parameter values
usually considered when modeling the accreting black hole sources, i.e.\
the emission originating from near above the ISCO.

We also constructed the normalized plots, where $g_{\rm{}max}(a)$
on the ordinate is divided by its value for $a=0.999$. These graphs are
given in the right panels of Figures \ref{fig2}--\ref{fig3} for
comparison with the unnormalized graph  for high inclinations
($\cos{i}\lesssim0.4$) in left panels. For lower inclinations we give
only normalized graphs ($\cos{i}\gtrsim0.5$; Figure~\ref{fig4}) because
in this case the dependence on the spin is very weak.

Finally, the normalized graphs are supplemented by Figure~\ref{fig5},
which has been constructed just for the fixed value of $a=0.999$. It
allows us to read the normalization factor and to reconstruct the
absolute values of the extremal shift in previous plots.

\section{Discussion}
\label{sec:discussion}
Wings of the relativistic line become more complex when the outgoing
signal is integrated over a finite range of radii. This
is also the case of the aggregate line profile that has been frequently
considered as originating from a radially extended zone of an accretion
disk. According to the relativistic
version of the standard disk model the emissivity has a maximum near
above the ISCO and it falls down towards the inner rim as well as
towards infinity, however, the dissipation in a hot corona does not need 
to follow this law. Therefore, the line radial emissivity cannot be
inferred solely from the standard disk model. Part of the information
from the spectral profile is lost in the radially integrated spectrum.

How could the extremal shifts be used to reconstruct, at least in
principle, the putative rings forming the spectral line radial
emissivity profile? The main underlying assumption requires that the
horns are resolved in the total observed profile. In fact, the right
panel of Figure~\ref{fig1} exhibits the individual components, which are
summed to form the final line profile. Each of these partial
constituents corresponds to one elementary ring,  radius of which can be
read from the $g_{\rm{}max}$ vs.\ $g_{\rm{}min}$ graph. In this way the
observed profile can be decomposed into the components. The required
time resolution of the method is of the order of orbital time at the
innermost ring.\footnote{Keplerian orbital time as function of radius
and spin of Kerr black hole is given by  $T_{\rm{}orb}(r;M,a) \doteq
310~\left(r^\frac{3}{2}+a\right) 
\frac{M}{10^7M_{\odot}}~\mbox{[sec]}.$} We further
assumed that an independent constraint on the disk inclination angle can
be given. This can be based for example on the ratio of equivalent
widths of the two horns. Then one will be able to read the emission
radius and the black hole spin from our graphs. Or, instead of the
graphical method, a fitting procedure can be employed using 
pre-computed tables of the energy shifts.

Naturally, this decomposition of an accretion disk into rings does  not
distinguish between the case of almost steady rings versus transient 
features that exist for a shorter period of time. The two cases should 
produce the same orbit-integrated profiles, so in this respect the 
assumption about the ring structure stands in the basis of our method.
This was discussed in more detail by \citet{czerny04}, who had developed
an approximation for the mean spectra of transient flares, which they
treat in terms of ``belts'' representing the time-averaged traces of the
flares on the disk surface. This scheme produces the ring structure of
the reflection spectra of the line emission fully consistent with the 
picture adopted in the current paper.

We note that another approach to the problem of constraining the radial
emissivity of the iron line, by well-resolved time-independent spectral
profiles, was discussed by \citet{cadez00}, or by using the hot-spot
scenario by \citet{murphy09}. However, the currently available data
do not allow us to achieve the high time resolution necessary to reveal the
individual orbiting spots in AGNs; this would require to study
time-scales of the order of $T_{\rm{}orb}$, which is for supermassive
black holes typically $\sim10^3$ sec and shorter. Therefore,
significantly higher collecting area is needed. Alternatively one
could apply this approach to accreting Galactic (stellar) black holes,
which can be brighter. In fact, the relativistic iron line has been
measured in several stellar black holes -- e.g.\ the case of GRS 1915+105 
microquasar \citep{martocchia02b}, or see the recent discussion of
XTE J1550-564 microquasar \citep[][and references cited
therein]{steiner10}. However, in the case of stellar-mass black holes
the time-scales are expected 
to be shorter, as follows from the mass-scaling relation.

The above-mentioned approaches offer a potentially interesting
application (though neither is useful in the context of present data).
Either significantly higher numbers of photon counts are required, or
one needs to catch the accretion disk in a state when a very small
number of well-separated annuli dominate the line emission, so that the
two different annuli of the accretion disk can be distinguished from
each other.

\section{Conclusions}
\label{sec:conclusions}
Calculation of the extremal energy shifts in
Section~\ref{sec:extremalshifts} and their graphical representation in
Figures \ref{fig2}--\ref{fig4} are the main results of the paper. The
principal assumption of the proposed application is about discrete rings
forming the line emission. Knowing the extremal shifts can be useful
also in another context, namely, the narrow (emission) lines produced by 
orbiting transient flares and spots on the accretion disk surface 
\citep{dovciak04a,demarco09}. Given the intrinsic emission energy, 
the extremal energy shifts define the range where these spectral 
features can appear in the observed spectrum.

We examined theoretical profiles of the relativistic spectral line
emerging from a set of concentric narrow rings which, as a whole, form a
radially extended zone of an inner accretion disk. In particular we
developed a systematical approach to determine the maximum and minimum
energy shifts of the observed line as a function of the model
parameters. As a motivation for our study we have mentioned
non-monotonic radial profiles of emissivity that are consistent with
intermittent episodes of accretion, and models of magnetized plasma
rings with radially periodic structure.

We constrained our calculations to photons arriving along direct light
rays, i.e., we ignored the higher-order images that could arise by rays
making several revolutions around the black hole. This constraint is
well-substantiated: firstly because the flux in these indirect images 
decreases exponentially with the image order ($n=2$, 3,\ldots), and
secondly these images are anyway blocked by the accretion disk. 

We also neglected some other complications,
such as the role of obscuration, the impact of geometrically thick 
(non-planar) shape of the accretion disk, or its warping. We expect 
that for example the role of source covering by intervening clouds along 
the line of sight \citep{karas00} will not affect the 
results although it could make the proposed method more difficult by 
enhancing the fluctuations of the observed signal. In case of AGN,
the role of accretion disk self-gravity can be important as it can 
significantly affects the vertical height of the outer regions of the 
disk \citep{karas04}. These effects should be considered in a future work.

\acknowledgments
We thank M.~Dov\v{c}iak, G.~Matt and T.~Pech\'a\v{c}ek for helpful
discussions. We acknowledge the Czech Science Foundation program (refs.\
205/07/0052 and 205/09/H033) and the European Space Agency PECS no.\
98040. The Astronomical Institute is supported by the Center for
Theoretical Astrophysics (LC06014).

\end{document}